\documentclass[aps,prl,twocolumn,showpacs,psfig]{revtex4}

\usepackage{amssymb}
\usepackage{amsmath}
\usepackage{graphicx}

\begin{document}

\title{The Active Traveling Wave in the Cochlea}
\author{Thomas Duke$^1$ and Frank J\"ulicher$^{2,3}$}
\affiliation{$^1$ Cavendish Laboratory, Madingley Road, Cambridge CB3 0HE, UK}
\affiliation{$^2$Institut Curie, Physicochimie, UMR CNRS/IC 168, 26 rue
d'Ulm, 75248 Paris Cedex 05, France}
\affiliation{$^3$ Max-Planck Institut f\"ur Physik komplexer Systeme,
N\"othnitzerstr. 38, 01187 Dresden, Germany}

\begin{abstract}
A sound stimulus entering the inner ear
excites a deformation of the basilar membrane
which travels along the cochlea towards the apex.
It is well established that this wave-like disturbance is amplified
by an active system.
Recently, it has been proposed that the active system consists
of a set of self-tuned critical oscillators which automatically
operate at an oscillatory instability. Here, we show how the
concepts of a traveling wave and of self-tuned critical oscillators can be
combined to describe the nonlinear wave in the cochlea.
\end{abstract}

\maketitle

\noindent
The sounds that we hear are detected within the cochlea of the inner
ear, a fluid-filled duct which is coiled like the chamber of a
snail's shell. This compartment is partitioned along almost its entire
length by the basilar membrane (BM). Incoming sound waves set the BM into
motion and its minute vibrations are monitored by specialized sensory
hair cells \cite{dall96}. The pioneering experiments of von B\'ek\'esy
\cite{beke60}, which were conducted on cadavers, demonstrated that sound
excites a traveling wave on the
BM, whose amplitude reaches a peak at a place which depends on the
frequency. This suggested that the cochlea acts as a spatial frequency
analyzer. When it became feasible to measure the BM response of living
specimens, a marked difference from von B\'ek\'esy's results was
revealed. The sharpness of filtering was greatly enhanced and the
response displayed pronounced nonlinear behavior close to
resonance \cite{rhod71,rugg97,robl01,russ97}.
These observations, together with the discovery that ears spontaneously
emit sounds at specific frequencies \cite{kemp79},
provided direct evidence for an active nonlinear amplifier in the cochlea
\cite{dall96}, as had
been foreseen by Gold \cite{gold48}.
But just how the ear could reliably employ an active
process without suffering from unstable behavior
has long been a matter of concern.

An active amplifier
which overcomes this
difficulty has recently been outlined
\cite{choe98,cama00,egui00}.
It has been proposed that the cochlea contains a set of
dynamical systems, each of which is maintained at the threshold of
an oscillatory instability by a self-tuning mechanism. Poised at
this critical point, on the verge of vibrating, each system is especially
responsive to periodic stimuli at its own characteristic frequency. The
concept of self-tuned critical oscillators \cite{cama00} can
account for the main features of hearing: sharp frequency
selectivity, extreme sensitivity and wide dynamic range; and also for
interference effects such as two-tone suppression and the generation of
combination tones \cite{juli01}. In this letter, we marry the concept of
critical oscillators with the physics of the traveling wave to provide a
unifying description of active cochlear mechanics.

{\it Cochlear waves.}---The basic physics of cochlear waves may be
described most succinctly by a one-dimensional model
\cite{zwis48,zwei76,debo80,ligh81,zwei91}. The BM separates the cochlear
duct
into two channels which are connected at the apex by a small aperture, the
helicotrema. A sound stimulus impinging on the oval window, at the base
of the cochlea, causes changes in the pressures
$P_1(x,t)$ and $P_2(x,t)$ in both channels. Here $t$ is the time and $x$
is the position along the cochlea, with the oval window at $x=0$ and the
helicotrema at $x=L$. The pressure gradients induce longitudinal currents
$J_1(x,t)$ and
$J_2(x,t)$, which flow in opposite directions in the two channels. We
define the relative current $j\equiv J_{1}-J_{2}$ and the
pressure difference
$p\equiv P_{1}-P_{2}$. Then the balance of pressure gradients and
inertial forces in the fluid may be written
\begin{equation}
        \rho \partial_{t}j=-bl\partial_{x}p \quad , \label{eq:jp}
\end{equation}
where $\rho$ is the fluid mass density, $l$ is the height of each
channel, and
$b$ is the breadth of the BM. The conservation of fluid volume implies
that a variation in the current along the cochlea must be accommodated by
a movement of the cochlear partition. We describe such deformations of the
BM by its height $h(x,t)$ as a function of time and position. Then
the conservation law is
\begin{equation}
     2 b \partial_{t} h +\partial_{x} j=0 \quad . \label{eq:cl}
\end{equation}
Combining this with Eq.~(\ref{eq:jp}), we obtain an equation for the BM
acceleration
\begin{equation}
        2 \rho b \partial_{t}^2 h =\partial_{x}\left[ bl \partial_{x}
        p\right ] \quad . \label{eq:hp}
\end{equation}
The pressure difference $p$ acts to deform the BM.
If the response is passive (eg. in the dead cochlea), the response
relation
close to the basal
end, where the stiffness $K(x)$ of the BM is high,
takes the simple form
\begin{equation}
       p(x,t) = K(x) h(x,t) \quad , \label{eq:ph}
\end{equation}
for small disturbances.
Eqs. (\ref{eq:hp}) \& (\ref{eq:ph}) together yield a linear wave
equation for the pressure, with local wave propagation velocity
\begin{equation}
        c(x) = \left(\frac{K(x) l}{2 \rho}\right)^{1/2} \quad .
\end{equation}

{\it Critical oscillators.}---In the active cochlea, the passive response
is
amplified by a force-generating system. This
system comprises a set of mechanical oscillators which are supported on
the BM, and which are positioned in such a way that they can drive
its motion. The characteristic frequency $\omega_{r}(x)$ of the
oscillators is a function of position along the membrane. In general,
such oscillators
could either vibrate spontaneously and thus generate
motion in the absence of a stimulus, or they could be quiescent and
behave like
a passive system. A particularly interesting case arises
at the boundary of these two regimes,
when every oscillator operates exactly at the critical
point where it undergoes an oscillatory instability.
Automatic regulation to this critical point ---
or Hopf bifurcation --- can in general be achieved by using a robust
self-tuning
mechanism based on
local feedback control \cite{cama00}. If the BM contains such self-tuned
critical oscillators, its deformation $h$ in response to pressure
differences
across the membrane $p$ has characteristic properties as a function of
frequency and amplitude, and nonlinear amplification occurs.

In order to describe this system, we first consider an individual
oscillator.
Its characteristic response to
periodic forcing at frequency $\omega$ can be written in a general form as
\cite{cama00}
\begin{equation}
\tilde p = A(\omega) \tilde h + B \vert \tilde h\vert^2 \tilde
h \quad . \label{eq:nl}
\end{equation}
Here, $\tilde h$ and $\tilde p$ are the Fourier amplitudes at
the forcing frequency and $A$ and $B$ are complex coefficients.
This expression follows from a systematic expansion in the oscillation
amplitude
$\tilde h$ which
is valid close to the Hopf bifurcation (comparable to
a Landau expansion of the free energy of thermodynamic systems near a
critical point).
Proximity
to an oscillatory instability thus automatically
provides for generic nonlinearities.
The dominant nonlinearity is cubic, a result that follows from
time-translation invariance.
The linear response coefficient
$A$ vanishes at the characteristic frequency
$\omega_{r}$ of the oscillator
so that,
at this particular frequency, the response
becomes purely nonlinear for small amplitudes.

Thus if we focus on a particular location
$x$ of the BM, its
response displays a nonlinear resonance when the frequency of the
stimulus is equal to the local characteristic frequency $\omega_{r}(x)$ of
the oscillators. The shape of the resonance, for
nearby frequencies, can be described by expanding the
function $A(\omega)$ in powers of $\omega-\omega_{r}(x)$. For
frequencies that differ substantially from
the local characteristic frequency,
on the other hand, we expect the active system to contribute little
to the BM response. In particular, when $\omega = 0$, the BM deflection is
governed only by its passive stiffness, according to Eq.~(\ref{eq:ph}).
We now assert that the simple functional form
\begin{equation}
A(x,\omega)=\alpha (\omega_{r}(x)-\omega) \quad , \label{eq:A}
\end{equation}
where $\alpha$ is a real constant, captures the essential
features of this BM response. Clearly it satisfies the requirement that
the linear response coefficient at location $x$ can be expanded about
$\omega_{r}(x)$. Secondly, it indicates that the passive stiffness is
proportional to the characteristic frequency: $K(x) = A(x,0) = \alpha
\omega_{r}(x)$. This corresponds well with experimental data. The
frequency-place map and the elasticity of the BM have been carefully
measured. Characteristic frequency and stiffness both decrease
approximately exponentially with distance along the cochlea, falling by
about two orders of magnitude from base to apex \cite{beke60,gree90}. We
therefore supplement Eqs. (\ref{eq:nl}) \& (\ref{eq:A}) with
\begin{equation}
      \omega_{r}(x)=\omega_{0} e^{-x/d} \quad , \label{eq:freqx}
\end{equation}
to obtain the full position-dependent response of the BM.
We take the coefficient $B$, describing the nonlinearity close to
resonance, to be a purely imaginary constant,
$B=i\beta$.  This simple choice ensures that Eq.~(\ref{eq:nl}) has
no spontaneously oscillating solution for $\tilde p=0$, as
required at the critical point.

{\it Active traveling waves.}---Combining Eq.~(\ref{eq:hp}) for the BM
acceleration with the response of an active membrane, described by
Eq.~(\ref{eq:nl}), we obtain a nonlinear wave
equation for the BM deformation.
In frequency representation, it reads
\begin{equation}
      -2 \rho b \omega^2 \tilde h = \partial_{x}\left
      [ bl \partial_{x} \left (A(x,\omega)\tilde h+ B\vert \tilde
      h\vert^2 \tilde h \right )\right ] \quad . \label{eq:hnlin}
\end{equation}
The complex solutions of this equation $\tilde h(x)=H(x) e^{i\phi(x)}$
describe the amplitude $H$ and the phase $\phi$ of the BM displacement
elicited by a periodic stimulus with incoming sound pressure
$p(x=0,t)=\tilde p(0)  e^{i\omega t}$.

For small pressures, the nonlinearity described by the cubic term in
Eq.~(\ref{eq:hnlin}) is significant only close to the resonant place
which, inverting Eq.~(\ref{eq:freqx}), is $x_{r}=d
\ln(\omega_{0}/\omega)$. Far
from this characteristic place, we obtain a linear wave equation which
can be solved in the  WKB approximation \cite{zwei76,ligh81}. The
approximate solution can be expressed as
\begin{equation}
     \tilde h(x) \sim \frac{1}{(\omega_{r}(x) -\omega)^{3/4}} \exp\left\{
     i\int_{0}^x dx'\; q(x')\right\} \quad , \label{eq:hlin}
\end{equation}
with local wave vector
\begin{equation}
      q(x)  =\left(\frac{2\rho}{l \alpha}\right)^{1/2}\frac{\omega}
      {\left({\omega_{r}(x)-\omega}\right)^{1/2}} \quad .
\end{equation}
At the basal end of the cochlea, $x<x_{r}$, $q$
is real and the solution is a traveling wave with a position-dependent
wave vector. As the wave propagates, its wavelength diminishes and its
amplitude builds up, until it approaches the place of resonance. In the
immediate vicinity of the characteristic place, $A$ decreases according to
Eq.~({\ref{eq:A}) while $\tilde h$ increases. Thus the cubic term in
Eq.~(\ref{eq:hnlin}) rapidly becomes more important than the linear
term. This
cuts off the divergence in Eq.~(\ref{eq:hlin}) and
leads to a strongly nonlinear BM response.
The wave peaks at
$x=x_p<x_r$, where the response displays the
characteristic nonlinearity of critical
oscillators,
$\tilde h(x_p)\sim \tilde p(x_p)^{1/3}$ \cite{cama00,egui00}.
From Eq. (\ref{eq:hlin}) we find that
$\tilde p(x_p)\sim A(x_p)^{1/4} \tilde p(0)$,
while the crossover from linear to nonlinear response
implies that $A(x_p)\sim  \vert B\vert\vert \tilde h(x_p)\vert^2$.
We thus find that the peak amplitude has a power law response
\begin{equation}
h(x_p)\sim p(0)^{\nu}
\end{equation}
as a function of the stimulus pressure at the base, with an exponent
$\nu=0.4$.
At positions beyond the characteristic
place, $x>x_{r}$, the wave vector $q$ becomes imaginary, indicating the
breakdown of wave propagation. The BM displacement decays very sharply in
this regime.

{\it Numerical solutions.}---Full solutions to the nonlinear wave
equation, Eq.~(\ref{eq:hnlin}),  can be obtained numerically. It is most
convenient to solve the equation for the pressure $\tilde p(x)$ which
satisfies
\begin{equation}
     -\mu \tilde p =(A+B u(\tilde p))\partial_{x}^2 \tilde p \quad ,
     \label{eq:pnlin}
\end{equation}
with $\mu=2\rho\omega^2/l$ and where
we have assumed for simplicity that $b$ and $l$ are constant
along the cochlea.
The variable $u=H^2$ is the squared deformation amplitude and is a
nonlinear function of $\tilde p$. Indeed, it follows from
Eq.~(\ref{eq:nl}) that $u(\tilde p)$ is the
unique real and positive root of the cubic equation
\begin{equation}
       \vert \tilde p\vert^2 = \vert A\vert^2 u +(A^* B + A B^*)u^2
       + \vert B\vert^2 u^3 \quad .
\end{equation}
Eq.~(\ref{eq:pnlin}) for $\tilde
p$ can be solved,
starting from $x=L$ and integrating towards
$x=0$. As a boundary condition, we impose zero pressure difference
at the helicotrema, $\tilde p(L)=0$, because the two cochlear channels are
connected there. A second boundary condition specifies the
value of
$\partial_{x}\tilde p$ at $x=L$. By varying this pressure gradient at
the helicotrema, we find solutions that correspond
to waves entering the cochlea at $x=0$ with different pressure amplitudes
$\tilde p(0)$. The profile of BM displacements can then be obtained from
the solution $\tilde p(x)$ via
\begin{equation}
\tilde h = \frac{\tilde p}{A+B u(\tilde p)} \quad .
\end{equation}

{\it Basilar membrane response.}---Examples of traveling waves are
displayed in Fig.\,1 for two different sound levels and varying stimulus
frequencies. Waves initiated at $x=0$ propagate with growing amplitude and
decreasing wavelength until they reach a point of resonance, beyond which
they decay rapidly. At 40 dB SPL, the resonance is sharp and the peak
response occurs at a location very close to the characteristic place
$x=x_{r}$, where the frequency of the active oscillators is equal to the
stimulus frequency. At 80 dB SPL, the resonance is much broader and the
location $x=x_p$
of maximal response shifts towards the base, in agreement with
experimental observations \cite{russ97}.

\begin{figure}[t]
\includegraphics[width=.60\linewidth]{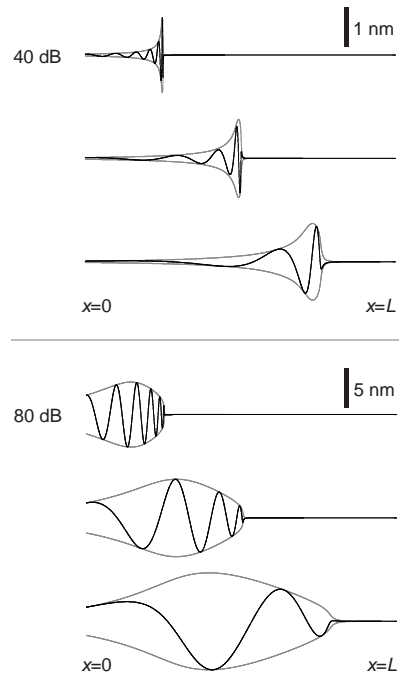}
 \caption{The active traveling wave on the BM for three
different frequencies ($f = 370\,{\rm Hz}, 1.3\,{\rm kHz}$ \& $4.6\,{\rm
kHz}$) whose corresponding characteristic places are
$x_r/L$ = 0.25 (top), 0.5 (center) \& 0.75 (bottom).
Instantaneous BM displacement
$h(x,t)$ (black lines) and wave amplitude $H(x)$ (gray lines) are shown
for two stimulus amplitudes, characterized by sound pressure level (SPL). At
40 dB SPL, a sharp resonance occurs at the characteristic place, where the
critical oscillators actively drive the response.  The resonance is
broader
at 80 dB SPL, and the peak shifts towards the base of the cochlea.
{\it Parameter values:} Cochlear dimensions $L = 35\,{\rm
mm}$ and $l = 1\,{\rm mm}$; fluid density $\rho = 10^3\,{\rm
kg/m^3}$; parameters governing the frequencies of the active
oscillators $\omega_0 = 10^5\,{\rm s^{-1}}$ and $d  = 7\,{\rm mm}$,
providing a frequency range of 100\,Hz--16\,kHz.
We choose $\alpha = 10^4\,{\rm Pa/ms}$, which implies a volumetric
stiffness of the BM varying in the range $6\times 10^6$--
$10^9\,{\rm Pa/m}$.
There is only one free parameter in our
calculations, $\beta = 10^{23}\,{\rm Pa/m^3}$, which we choose to
fit the nonlinearity of the response according to
sound pressure level (SPL). For simplicity, it is assumed
that the middle ear raises sound pressures by 20\,dB, independent of
frequency.}  
\end{figure}

The response at a particular location on the BM exhibits the qualitative
properties of cochlear tuning that have been observed experimentally
\cite{rhod71,rugg97,robl01}. Fig.\,2a displays the gain
$\omega\vert\tilde h\vert/\vert \tilde p(0)\vert$ of BM
velocity, obtained from our numerical solutions, as a function of stimulus
frequency for different sound levels. At low frequencies the response is
linear and the gain is independent of the stimulus amplitude. As the
stimulus frequency approaches the resonant frequency, the response becomes
nonlinear and the gain diverges as the SPL declines. At higher frequencies,
the response drops precipitously. The magnitude of the BM displacement,
shown in Fig.\,2b,
is typically several nanometers at resonance, in quantitative agreement with
experimental data \cite{rugg97}.  The phase $\phi$ of the traveling wave at
a particular
location on the BM is displayed in Fig.\,2c. As observed experimentally, it
decreases with
increasing frequency --- gradually at first, but more abruptly as resonance
is approached
--- and then varies only little at frequencies higher than the
characteristic frequency.

\begin{figure}[t]
\includegraphics[width=.70\linewidth]{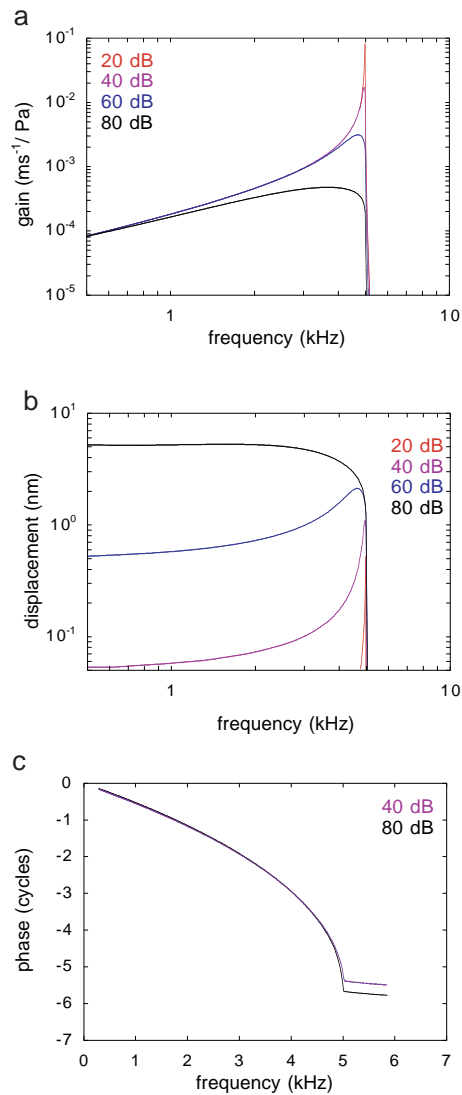}
\caption{Response of the BM at a fixed position as a function
of stimulus frequency, for different sound pressures. The characteristic
frequency of the active oscillators at that location is 5\,kHz. (a)
Velocity gain, i.e. BM velocity divided by sound pressure of the
stimulus.
(b) Corresponding BM displacement.
(c) Phase difference between
stimulus and BM oscillation.}
\end{figure}

{\it Discussion.}---In many recent cochlear models,
sections of the BM are considered to behave as
inertial oscillators which are either lightly damped (in
passive models) \cite{zwei76,debo80,ligh81} or driven by internal
forces (in active models) \cite{duif85,kols90,debo96}.
The characteristic frequency at a particular location is then the local
resonant frequency
of the BM, which varies as the square root of the stiffness. A
problem with this interpretation is that, in order to obtain
the observed range of characteristic frequencies, the stiffness of the BM
would have to vary by more than four orders of magnitude from base to
apex. The measured variation is only a factor of one hundred
\cite{beke60,naid98}. This difficulty is circumvented by our theory,
where the range of frequencies at which the BM resonates
is determined by the frequencies of the oscillators that are ranged
along it, and is not governed by the stiffness or the inertia.
Some models of the active cochlea are very specific and rely on additional
degrees of freedom, secondary resonances or time-delayed feedbacks
\cite{debo96}. Such
descriptions lack the simplicity and generality of our
approach and miss the generic nature of the power-law nonlinearities
\cite{cama00,egui00,magn_pp} conferred by the Hopf bifurcation.

In this letter it has been our aim to provide a concise, coherent
interpretation of a wide variety of observations, rather than a detailed
fit of individual data. Considering that our model incorporates only one
free parameter whose value is not determined by independent measurement,
the qualitative agreement with a diverse set of experimental data is
striking. We have not sought to specify the physical nature of the
active oscillators. The electromotility of outer hair cells has been
implicated in active amplification in the mammalian cochlea
\cite{brow85,ashm87,dall96}, but the motile response of hair bundles
may also play a role. Indeed, the hair bundles of frog
hair cells have recently been demonstrated to behave as
Hopf oscillators \cite{mart01a,mart01b}. Because the response of
self-tuned critical oscillators is generic, our analysis remains valid
whatever the physical basis of force generation.

We thank M. Magnasco, P. Martin,
E. Olson, J. Prost and G. Zweig for stimulating discussions. T.D. is a
Royal Society University Research Fellow.

\end{document}